\documentclass[aps,prb,preprint,groupedaddress,showpacs,amsmath,amssymb]{revtex4}

\usepackage{graphicx}% Include figure files
\usepackage{dcolumn}% Align table columns on decimal point
\usepackage{bm}% bold math

\begin{document}

%Title of paper
\title{Anisotropic thermodynamic and transport properties of single crystalline (Ba$_{1-x}$K$_x$)Fe$_2$As$_2$ ($x = 0$ and $0.45$).}

% repeat the \author .. \affiliation  etc. as needed
% \email, \thanks, \homepage, \altaffiliation all apply to the current
% author. Explanatory text should go in the []'s, actual e-mail
% address or url should go in the {}'s for \email and \homepage.
% Please use the appropriate macro foreach each type of information

% \affiliation command applies to all authors since the last
% \affiliation command. The \affiliation command should follow the
% other information
% \affiliation can be followed by \email, \homepage, \thanks as well.
\author{N. Ni$^{1,2}$}
\author{S. L. Bud'ko$^{1,2}$}
\author{A. Kreyssig$^{1,2}$}
\author{S. Nandi$^{1,2}$}
\author{G. E. Rustan$^{1,2}$}
\author{A. I. Goldman$^{1,2}$}
\author{S. Gupta$^{1,3}$}
\author{J. D. Corbett$^{1,3}$}
\author{A. Kracher$^1$}
\author{P. C. Canfield$^{1,2}$}

\affiliation{$^1$Ames Laboratory US DOE, Iowa State University, Ames, Iowa 50011}
\affiliation{$^2$Department of Physics and Astronomy, Iowa State University, Ames, Iowa 50011}
\affiliation{$^3$Department of Chemistry, Iowa State University, Ames, Iowa 50011}

\date{\today}

\begin{abstract}
Single crystals of BaFe$_2$As$_2$ and (Ba$_{0.55}$K$_{0.45}$)Fe$_2$As$_2$ have been grown out of excess Sn with 1\% or less incorporation of solvent.  The crystals are exceptionally micaceous, are easily exfoliated and can have dimensions as large as 3 x 3 x 0.2 mm$^3$.   The BaFe$_2$As$_2$ single crystals manifest a structural phase transition from a high temperature tetragonal phase to a low temperature orthorhombic phase near 85 K and do not show any sign of superconductivity down to 1.8 K.  This transition can be detected in the electrical resistivity, Hall resistivity, specific heat and the anisotropic magnetic susceptibility.  In the (Ba$_{0.55}$K$_{0.45}$)Fe$_2$As$_2$ single crystals this transition is suppressed and instead superconductivity occurs with a transition temperature near 30 K.  Whereas the superconducting transition is easily detected in resistivity and magnetization measurements, the change in specific heat near $T_c$ is small, but resolvable, giving $\Delta C_p/\gamma T_c \approx 1$.  The application of a 140 kOe magnetic field suppresses $T_c$ by only $\sim 4$ K when applied along the c-axis and by $\sim 2$ K when applied perpendicular to the $c$-axis.  The ratio of the anisotropic upper critical fields, $\gamma = H_{c2}^{\perp c} / H_{c2}^{\| c}$, varies between 2.5 and 3.5 for temperatures down to $\sim 2$ K below $T_c$.
\end{abstract}

% insert suggested PACS numbers in braces on next line
% insert suggested keywords - APS authors don't need to do this
%\keywords{}

%\maketitle must follow title, authors, abstract, \pacs, and \keywords
\maketitle

\section{Introduction}
The recent discovery of superconductivity in F-doped LaFeAsO near 25 K \cite{jap_dis} along with the ability to essentially double $T_c$ either by application of hydrostatic pressure \cite{jap_pre} or by substitution of rare earths \cite{fst_sm} has given rise to a new sense of optimism within the high temperature superconductivity community.  First of all, these materials break the tyranny of copper, offering high transition temperatures without requiring the presence of copper oxide.  Secondly it is not at all clear (experimentally) what the mechanism for superconductivity in these materials is, although there are a multitude of exotic and intriguing theoretical predictions. Unfortunately, the growth of single crystals of these compounds, specifically the superconducting variants, has proven to be difficult.  Single grains as large as 300 $\mu$m have been characterized as part of polycrystalline pellets (synthesized at 3.3 GPa) \cite{pro_po} and free standing crystallites $\sim$ 100 $\mu$m across have been grown out of high temperature salt solutions (again at high pressures).  \cite{karp}

A second, key finding has been the even more recent discovery of superconductivity below $T_c = 38$ K in (Ba$_{1-x}$K$_x$)Fe$_2$As$_2$. \cite{sec_ger}  Whereas BaFe$_2$As$_2$, like LaFeAsO, is a tetragonal compound with a Fe plane capped above and below by As, \cite{fst_ger}  it does not contain oxygen.  This means that superconductivity in the (Ba$_{1-x}$K$_x$)Fe$_2$As$_2$ system not only breaks the tyranny of copper, it also implies that the superconductivity of these FeAs-based compounds is not intimately associated with oxide physics.  In addition to this conceptual simplification, there is a profound experimental simplification as well:  BaFe$_2$As$_2$ is an intermetallic compound and can be grown in single crystal form utilizing conventional growth from a metal solvent, or flux.  In this paper we present crystallographic, as well as anisotropic thermodynamic and transport data on single crystals of the non-superconducting parent compound, BaFe$_2$As$_2$ as well as on superconducting (Ba$_{0.55}$K$_{0.45}$)Fe$_2$As$_2$.

\section{Experimental Methods}
Single crystals of BaFe$_2$As$_2$ and (Ba$_{0.55}$K$_{0.45}$)Fe$_2$As$_2$ were grown out of a Sn flux, using conventional high temperature solution growth techniques. \cite{can_fi}  Elemental Ba, K, Fe and As were added to Sn in the ratio of [(Ba/K)Fe$_2$As$_2$] : Sn  =  1 : 48 and placed in an $\sim$ 2 ml MgO crucible.  A second, catch, crucible containing silica wool was placed on top of the growth crucible and both were sealed in a silica ampoule under approximately 1/3 atmosphere of argon gas.  It should be noted that the packing and assembly of the growth ampoule was performed in a glove box with a nitrogen atmosphere.  The sealed ampoule was placed in a programmable furnace and heated to either 1000 $^\circ$C or 850 $^\circ$C and cooled over 36 hours to 500 $^\circ$C.  Once the furnace reached 500 $^\circ$C the Sn was decanted from the (Ba$_{1-x}$K$_x$)Fe$_2$As$_2$  crystals.  Whereas the BaFe$_2$As$_2$ growth was cooled from 1000 $^\circ$C, when the potassium doped growths were taken to 1000 $^\circ$C there was a clear attack on the silica wool as well as on the silica tube and a greatly reduced amount of K in the resulting crystals.  In order to avoid this loss of K, $x > 0$ growths of (Ba$_{1-x}$K$_x$)Fe$_2$As$_2$ were only heated to 850 $^\circ$C.  Even with this precaution in place, we found it necessary to use excess K in the growth to compensate for losses due to the combination of potassium vapor pressure and reactivity.  The (Ba$_{0.55}$K$_{0.45}$)Fe$_2$As$_2$ growth used an initial stoichiometry of (Ba$_{0.6}$K$_{0.8}$)Fe$_2$As$_2$.

\begin{figure}
\begin{center}
\includegraphics[angle=0,width=120mm]{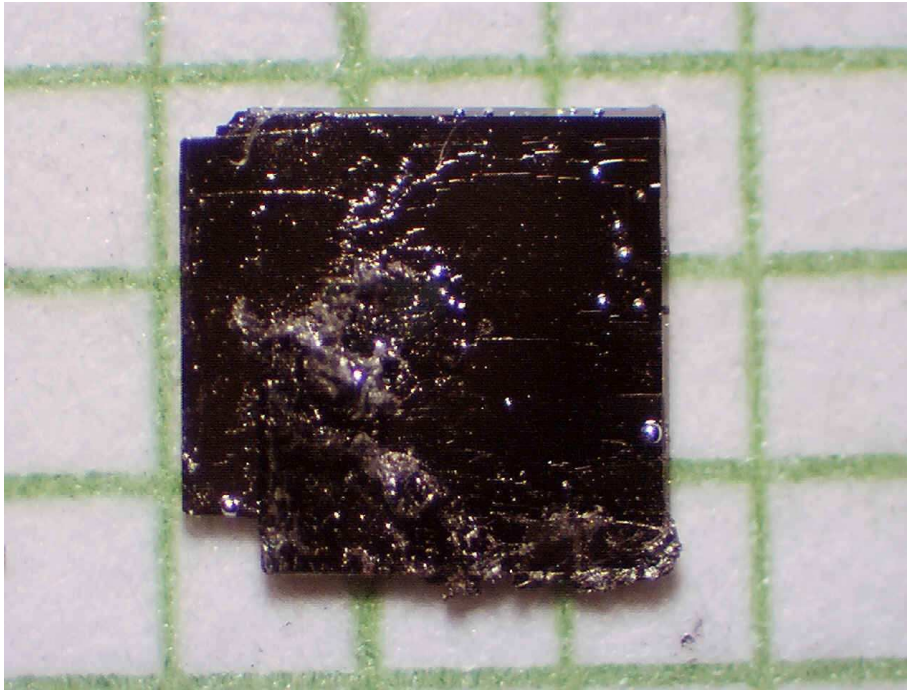}
\end{center}
\caption{(Color online) Photograph of a single crystal of (Ba$_{0.55}$K$_{0.45}$)Fe$_2$As$_2$ on a mm grid.  The crystallographic $c$-axis is perpendicular to the plane of the plate.  Droplets of Sn flux can be seen on the surface.}\label{F1}
\end{figure}

Figure \ref{F1} shows a picture of a single crystal of (Ba$_{0.55}$K$_{0.45}$)Fe$_2$As$_2$ against a mm scale.  Typical crystals have dimensions of 1-2 mm $\times$ 1-2 mm $\times$ 0.05-0.1 mm.  Some crystals manifest linear dimensions as large as 3-4 mm.  The crystallographic $c$-axis is perpendicular to the plane of the plate-like single crystals.  All of the crystals grown so far have been rather thin and are prone to exfoliation.

Powder X-ray diffraction data was collected on a Rigaku Miniflex diffractometer using Cu $K \alpha$ radiation at room temperature.  Room temperature, single crystal diffraction was performed on a Bruker SMART CCD four circle diffractometer and analyzed by standard methods using the  SHELXTL 6.1 software package.  Temperature dependent, single crystal X-ray diffraction measurements were performed on a standard, four-circle diffractometer using Cu $K \alpha$  radiation from a rotating anode X-ray source, selected by a germanium $(1~1~1)$ monochromator.  For the measurements, a plate like single crystal with dimensions of 1.5 $\times$ 1.5 $\times$ 0.2 mm$^3$ was selected and attached to a flat copper sample holder on the cold finger of a closed cycle, displex refrigerator.  The diffraction patterns were recorded while the temperature was varied between 10 K and 300 K.  The mosaicity of the investigated BaFe$_2$As$_2$ single crystal was $0.04 ^\circ$ FWHM in the rocking curve of the $(0~0~10)$ reflection.  Elemental analysis of the samples was performed using wavelength dispersive x-ray spectroscopy (WDS) in the electron probe microanalyzer (EPMA) of a JEOL JXA-8200 Superprobe.

Magnetic field and temperature dependent magnetization data were collected using a Quantum Design (QD) Magnetic Properties Measurement System (MPMS) and temperature dependent specific heat as well as magnetic field and temperature dependent electrical transport data were collected using a QD Physical Properties Measurement System (PPMS).  Electrical contact was made to the samples using Epotek H20E silver epoxy to attach Pt wires in a 4-probe configuration.  (It is worth noting that curing of the epoxy at 120 $^\circ$C for up to 30 minutes does not seem to degrade the samples.)  Basal plane resistivity values were determined by estimating the length and cross sections of the samples assuming that current density was uniformly distributed throughout the cross section.  Given the layered nature and the thinness of the crystals we believe that the resistivity values should be accurate to better than $\pm 50\%$.  Hall measurements on BaFe$_2$As$_2$ were performed in a QD PPMS instrument using four probe ac ($f = 16$ Hz, $I = 1$ mA) technique with current flowing in the $ab$ plane approximately parallel to the $a$-axis and field parallel to the $c$-axis. To eliminate the effect of a misalignment of the voltage contacts, the Hall measurements were taken for two opposite directions of the applied field, $H$ and $-H$, and the odd component, $[\rho_H(H) - \rho_H(-H)]/2$, was taken as the Hall resistivity.

\section{Results}
Given that the stoichiometry of these materials appears to play a key role in their physics, elemental analysis was performed on both the nominally pure BaFe$_2$As$_2$ sample as well as on the K-doped sample.  WDS measurements on the pure sample gave a Ba:Fe:As ratio of 1:2:2 with approximately 1\% Sn appearing to be present as well. (These results are similar to those shown below for the K-doped sample, but with slightly higher values of incorporated Sn.)  As will be discussed below, it is possible that this is not simply surface Sn associated with small amounts of residual flux, but rather Sn that has been incorporated into the BaFe$_2$As$_2$ structure.  Fortunately, if indeed there is a small amount of Sn incorporated into the structure, it does not prevent the stabilization of the superconducting state via K-doping.

\begin{figure}
\begin{center}
\includegraphics[angle=0,width=120mm]{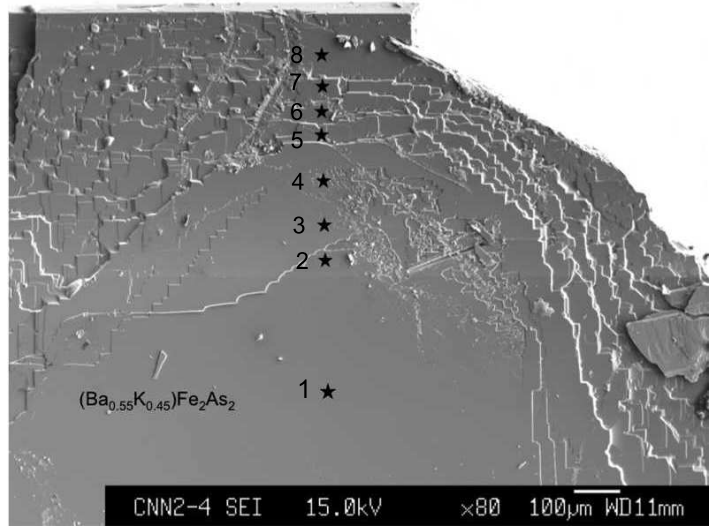}
\end{center}
\caption{Electron microscope image of plate of (Ba$_{0.55}$K$_{0.45}$)Fe$_2$As$_2$.  Elemental analysis was performed via WDS at each of the points marked by a star. The average K substitution value was found to be 0.44 with a 0.07 standard deviation.}\label{F2}
\end{figure}

The K-content of the doped sample was also determined via WDS analysis. Figure \ref{F2} shows an image of the sample measured as well as the location of the spots tested.  Table \ref{T1} presents the atomic percents of each element estimated from the weight percents determined from the analysis.  As can be seen the sample was neither polished nor cleaved.  Instead, different terraces on the as grown surface were measured. This terracing is a clear example of the layering so common to these crystals:  each of the visible layers can be easily cleaved or exfoliated from the rest of the sample. The data in Table \ref{T1} are consistent with a 1:2:2 ratio of (Ba + K):Fe:As.  These data are also consistent with an $\sim 0.66$\% (atomic) of Sn being incorporated into the sample (most likely on the As site).  The strong anti-correlation between the K and Ba values is consistent with K substituting on the Ba site.  As can be seen by comparing Figure \ref{F2} with Table \ref{T1}, there is no systematic variation of the K content from terrace to terrace.  Treating the sample as (Ba$_{1-x}$K$_x$)Fe$_2$As$_2$, the average $x$ value was found to be 0.45 with a standard deviation of  0.07.  For notational simplicity we will refer to these samples as $x = 0.45$, i.e. (Ba$_{0.55}$K$_{0.45}$)Fe$_2$As$_2$, but it should be kept in mind that there may be a degree of variation in the K content from layer to layer and, possibly, from crystal to crystal.

\begin{figure}
\begin{center}
\includegraphics[angle=0,width=120mm]{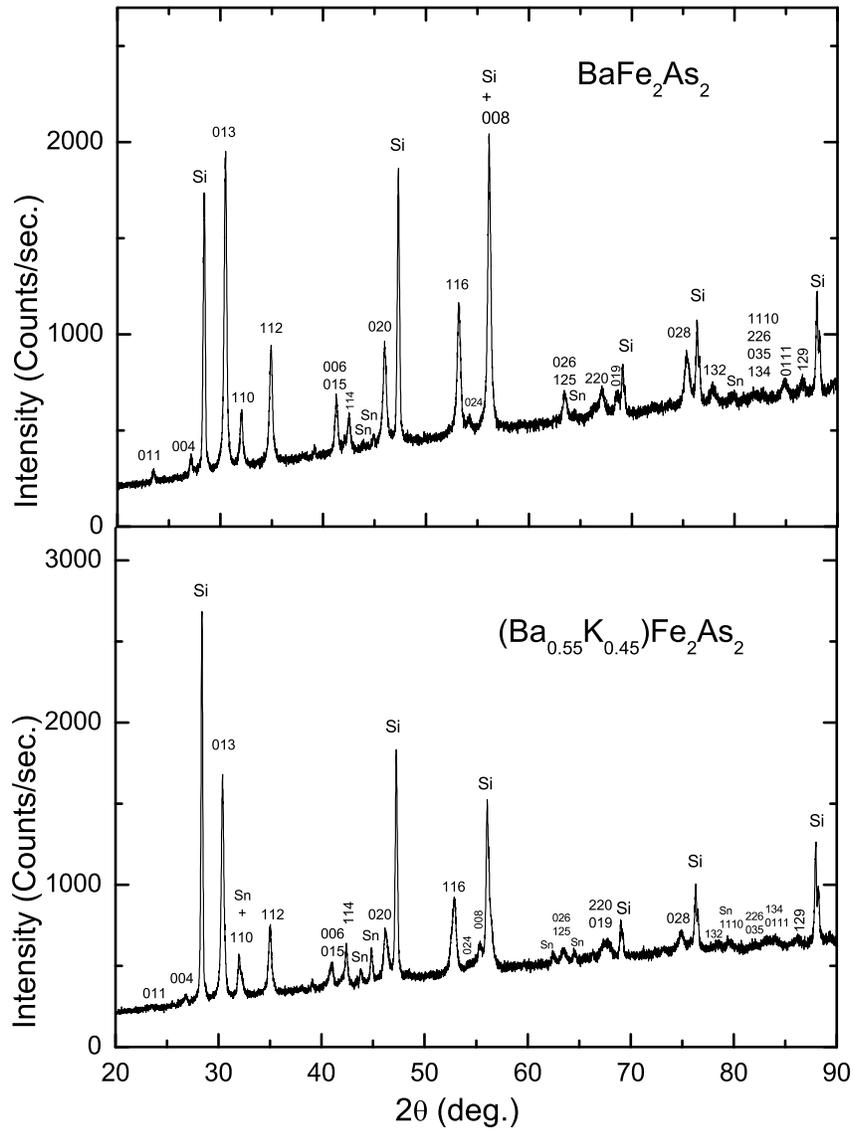}
\end{center}
\caption{Powder X-ray diffraction patterns from BaFe$_2$As$_2$ (upper panel) and (Ba$_{0.55}$K$_{0.45}$)Fe$_2$As$_2$ (lower panel).  Powdered Si was added as a standard and Sn reflections are present due to small amounts of residual flux on the surfaces of the crystal that were ground to provide the powder.}\label{F3}
\end{figure}

Powder x-ray diffraction patterns were taken on ground, single crystals from each batch.  The data are shown in Fig. \ref{F3}.  For BaFe$_2$As$_2$, the lattice parameters were found to be $a = 3.9436~\AA$ and $c = 13.115~\AA$.  Room temperature, single crystal diffraction data taken on a crystal of BaFe$_2$As$_2$ gave similar results:  $a = 3.9467~\AA$ and $c = 13.113~\AA$.  Whereas the $a$-lattice parameter is similar to that already reported \cite{fst_ger}, the $c$-lattice parameter is slightly, but significantly, larger.  This may be associated with the incorporation of Sn into the bulk, as discussed above.  Unfortunately, the current crystals did not allow for the collection of enough reflections to allow for a refinement of such a small, possible Sn occupancy.  The lattice parameters inferred for (Ba$_{0.55}$K$_{0.45}$)Fe$_2$As$_2$ from the powder x-ray diffraction data were $a = 3.9186~\AA$ and $c = 13.256~\AA$, comparable to the values found for the 40\% K substituted polycrystalline samples. \cite{sec_ger}

\begin{figure}
\begin{center}
\includegraphics[angle=0,width=120mm]{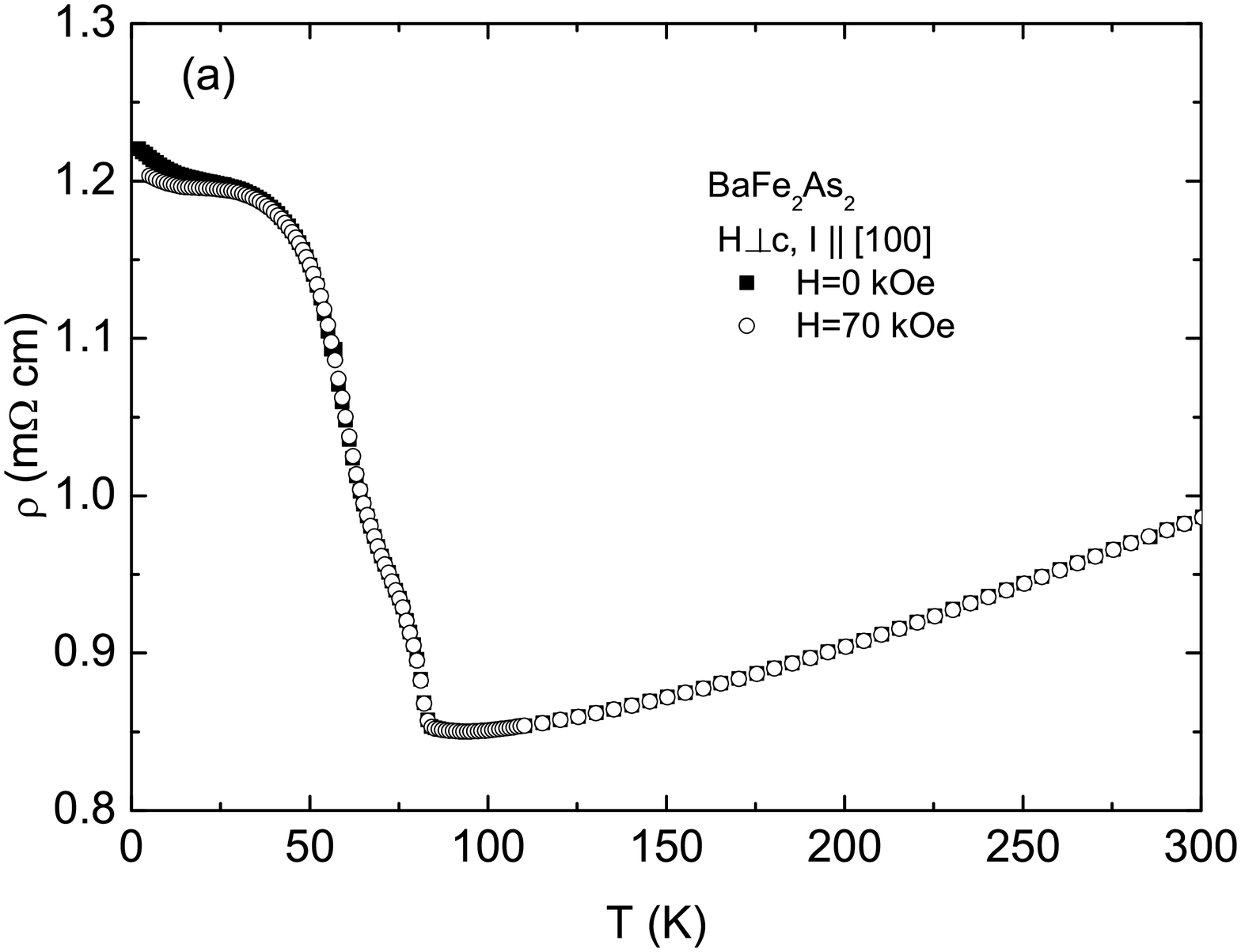}
\includegraphics[angle=0,width=120mm]{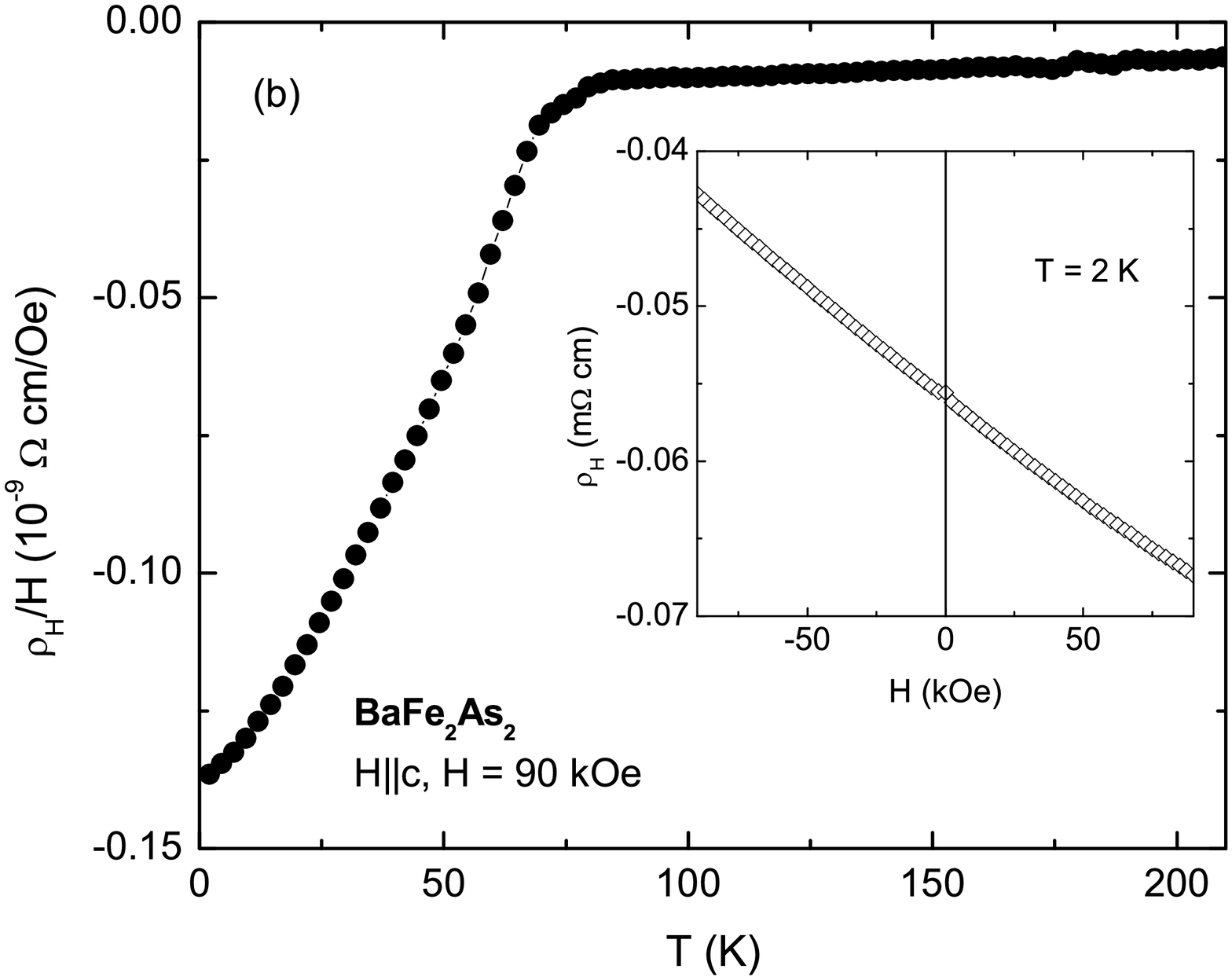}
\end{center}
\caption{(a) Temperature dependence of the in-plane electrical resistivity of BaFe$_2$As$_2$ for zero and 70 kOe applied magnetic field.  (b) Temperature dependence of the Hall resistivity divided by applied magnetic field.  The inset shows the linearity of the Hall resistivity at $T = 2$ K.}\label{F4}
\end{figure}

Figure \ref{F4}(a) presents temperature dependent electrical resistivity data for single crystalline BaFe$_2$As$_2$.  For 300 K$ > T > $90 K the resistivity is relatively large and weakly temperature dependent; as the sample is cooled below $\sim$ 85 K the resistivity increases rapidly, and at 4 K the resistivity is approximately 50\% larger than it was at 90 K.  The application of  a 70 kOe magnetic field perpendicular to the $c$-axis has no effect on the $T \sim 85$ K phase transition and only the smallest effect on the $T < 10$ K resistivity.  Similar behavior was seen for several samples from the same batch.  Figure \ref{F4}(b) shows the temperature dependent Hall resistivity divided by applied field ($\rho_H/H$) which also manifests a clear and marked increase in its absolute value as the sample is cooled below $\sim$ 85 K.  There is no evidence for superconductivity in BaFe$_2$As$_2$ for $T > 1.8$ K in either of these measurements.

\begin{figure}
\begin{center}
\includegraphics[angle=0,width=120mm]{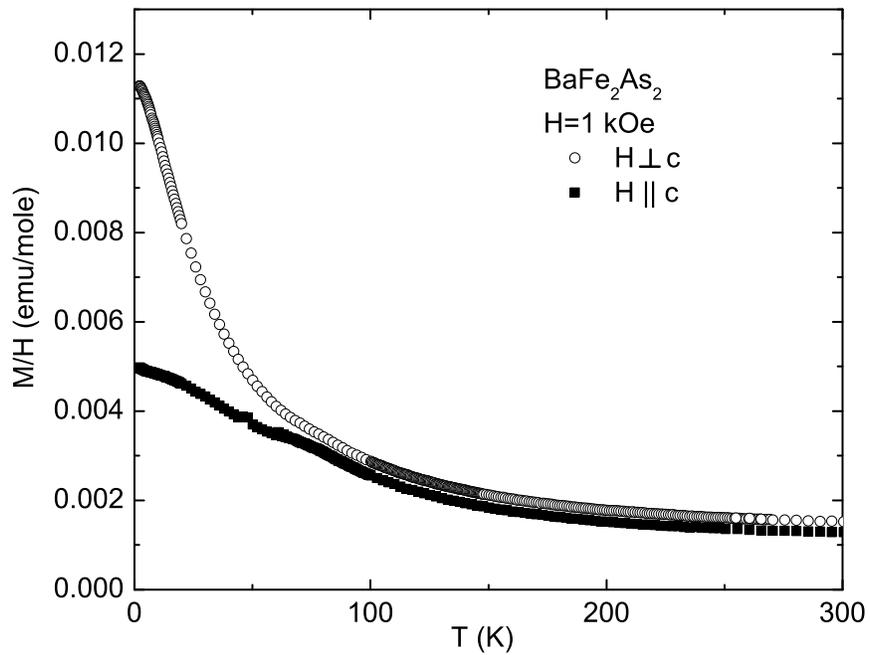}
\end{center}
\caption{Temperature dependent, anisotropic magnetization divided by applied field for BaFe$_2$As$_2$.}\label{F5}
\end{figure}

\begin{figure}
\begin{center}
\includegraphics[angle=0,width=120mm]{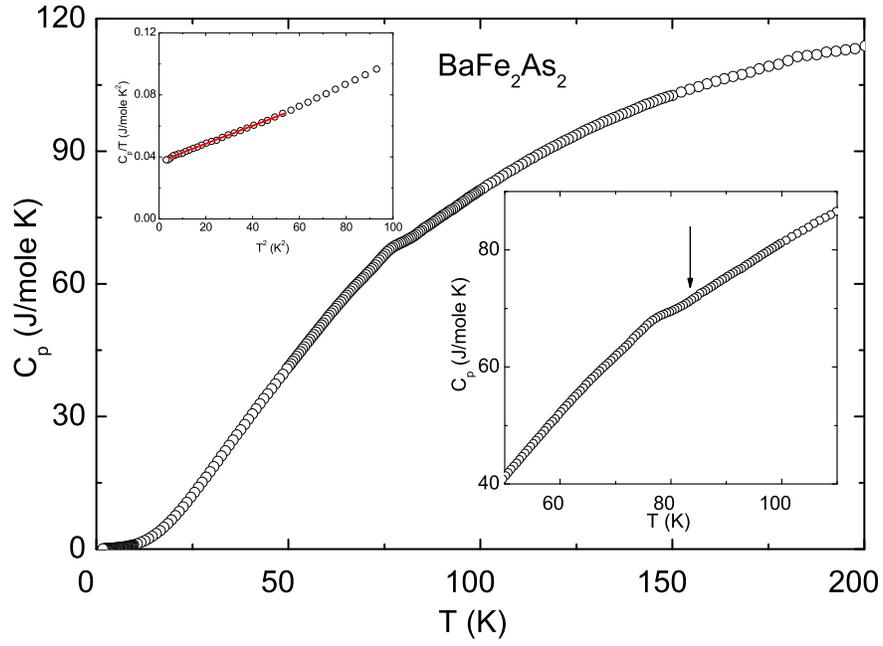}
\end{center}
\caption{(Color online) Temperature dependent specific heat of BaFe$_2$As$_2$.  Upper inset:  $C_p/T$ as a function of $T^2$ for low temperature data; lower inset: expanded view near the $T \sim 85$ K phase transition.}\label{F6}
\end{figure}

The anisotropic magnetic susceptibility data are presented in Fig. \ref{F5}.  Near room temperature the sample is close to isotropic, but below $\sim$ 85 K there is a clear anisotropy with $\chi_{\|c} < \chi_{\perp c}$.  Although the $T \sim 85$ K transition is more clearly seen in the $\chi_{\|c}$ data, there is a clear break in slope in the $\chi_{\perp c}$ data near this temperature as well.  The temperature dependent specific heat data for BaFe$_2$As$_2$ are shown in Fig. \ref{F6}.  There is a broad feature near 85 K and for $T^2 < 40$ K$^2$ the data can be fit to the standard power law, $C_p = \gamma T + \beta T^3$ with $\gamma = 37$ mJ/mole K$^2$ and $\beta = 0.60$ mJ/mole K$^4$.  The value for the Debye temperature, $\Theta_D$, inferred from this is 250 K.
	
The transport and thermodynamic data shown in Figs. \ref{F4},\ref{F5},\ref{F6} show a feature near 85 K and do not manifest any sign of the 140 K transition reported for the polycrystalline samples of BaFe$_2$As$_2$. \cite{fst_ger}  Given that these single crystals have been grown out of excess Sn, appear to have a small, but detectable amount of Sn incorporated into their bulk, and show a slightly expanded $c$-axis, it is plausible that the changes that we have found are another manifestation of the fact that this transition is very sensitive to small changes in stoichiometery, band filling and lattice parameter.  This being said, a valid question is, what is the nature of the 85 K transition in these crystals and is it similar to the structural phase transition that has already been found for in polycrystalline samples?
	
\begin{figure}
\begin{center}
\includegraphics[angle=0,width=90mm]{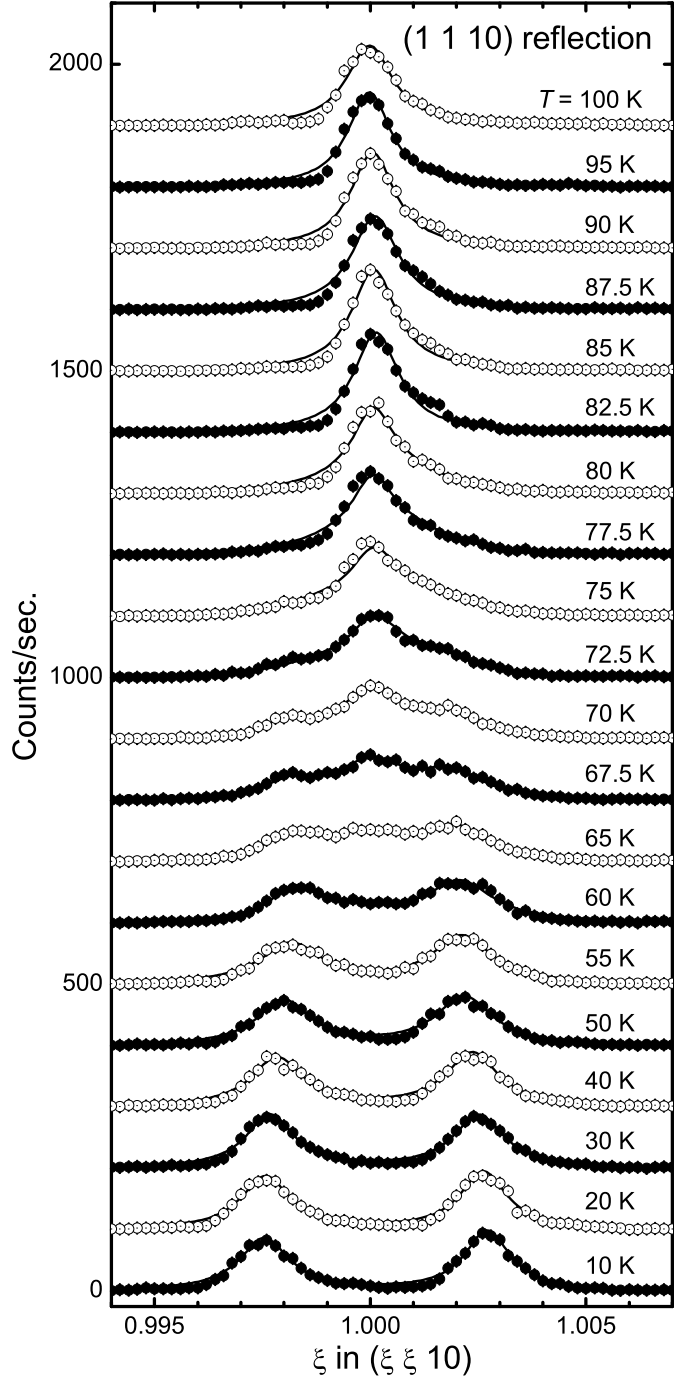}
\end{center}
\caption{X-ray diffraction scans along the $(1~1~0)$ direction through the position of the tetragonal $(1~1~10)$ reflection for selected temperatures.  The lines represent the fitted curves to obtain the reflection positions for the data shown in Fig. 8. The offset between each data set is 100 Counts/s.}\label{F7}
\end{figure}

\begin{figure}
\begin{center}
\includegraphics[angle=0,width=90mm]{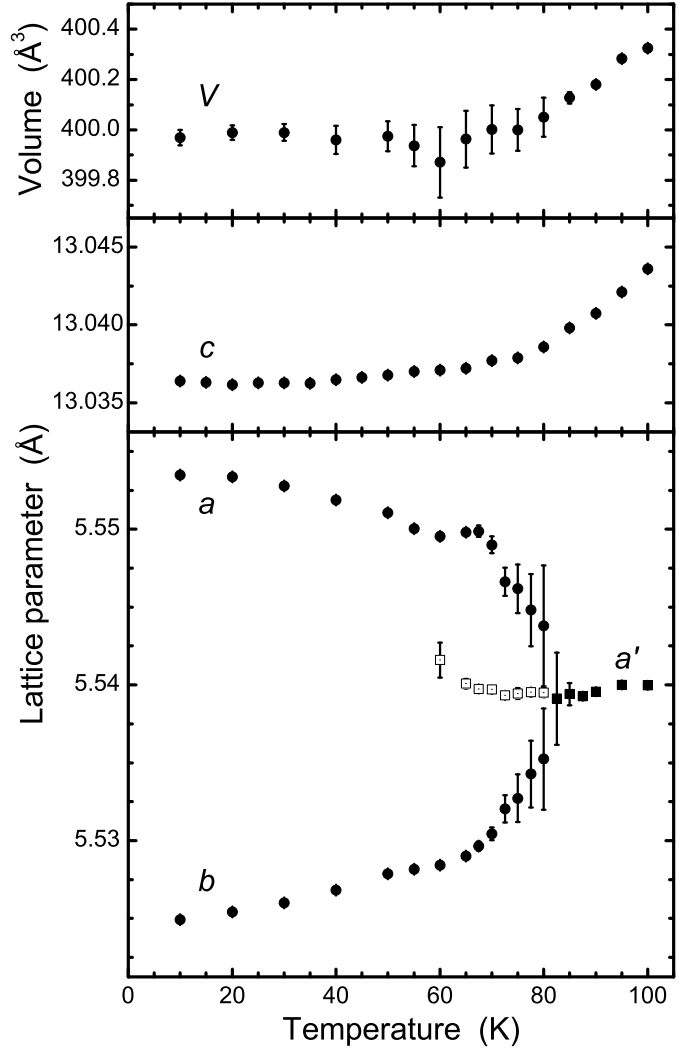}
\end{center}
\caption{Lattice parameters and unit cell volume for the tetragonal and orthorhombic phases as extracted from the data shown in Fig. 7 for the $(1~1~10)$ reflection and from the $(0~0~10)$ reflection (not shown).  The open squares in the lower panel represent the lattice parameter corresponding to the central reflection which coexists with the pair of reflections related to the orthorhombic phase (seen clearly in the 70 K data set in Fig. 7).  For clarity, the $a$-lattice parameter in the high-temperature, tetragonal phase has been multiplied by a factor of $\sqrt{2}$ so as to allow for comparison to the low temperature orthorhombic phase data.  The middle panel presents the temperature dependence of the $c$-lattice parameter and the upper panel presents the calculated unit cell volume.  The error bars represent the relative precision, the absolute error is
method-related and significantly larger.}\label{F8}
\end{figure}

To address this question, Figs. \ref{F7} and \ref{F8} summarize the temperature dependent, single crystal X-ray diffraction data collected on BaFe$_2$As$_2$.  Figure \ref{F7} shows the splitting of the $(1~1~10)$ reflection as the sample is cooled through $T \sim 85$ K.  Whereas there is splitting in the $(1~1~10)$ reflection in $(\xi~\xi~0)$ scans below 85 K there is no change in the shape of the $(0~0~10)$ reflection between 10 K and 160 K.  This is consistent with the reported tetragonal-to-orthorhombic phase transition, from space group $I4/mmm$ to $Fmmm$, with a distortion along the $(1~1~0)$ direction.  Figure \ref{F8} plots the temperature dependence of the lattice parameters between 10 and 100 K.  In addition to the splitting of the tetragonal $a$-lattice parameter below 85 K, the $c$-lattice parameter and the unit cell volume change their temperature dependencies below 85 K as well, becoming essentially temperature independent at lower temperatures.  The relative splitting of the lattice parameter in the orthorhombic phase is approximately 70\% of the splitting found in the polycrystalline material. \cite{fst_ger}
	
Although the tetragonal to orthorhombic splitting is very clear when comparing the 10 K and 100 K scans (Fig. \ref{F7}), in the intermediate temperature range (85 K to 60 K) reflections related to the orthorhombic phase appear to coexist with reflections related to the tetragonal phase.  This could imply a first order nature to this transition, but a more trivial explanation could be that the X-ray beam illuminated parts of the sample with slightly different transition temperatures.
	
\begin{figure}
\begin{center}
\includegraphics[angle=0,width=120mm]{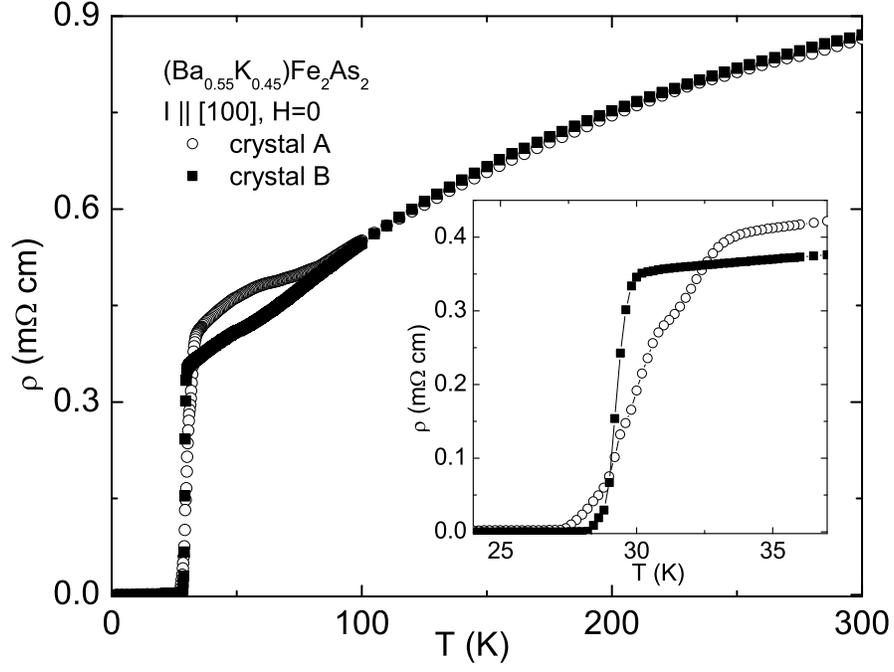}
\end{center}
\caption{Temperature dependence of the in-plane electrical resistivity of two crystals of (Ba$_{0.55}$K$_{0.45}$)Fe$_2$As$_2$.  Inset:  expanded scale for temperatures near the superconducting transition.}\label{F9}
\end{figure}

The temperature dependent electrical resistivity data for two separate single crystals from the same batch of (Ba$_{0.55}$K$_{0.45}$)Fe$_2$As$_2$ are presented in Fig. \ref{F9}.  Both crystals have $T > 100$ K resistivities that are very similar in size and temperature dependence to that found in Fig. \ref{F4} for BaFe$_2$As$_2$.  Both crystals also manifest superconductivity below $\sim 30$ K.  Crystal A shows a break in slope near the 85 K transition temperature found in BaFe$_2$As$_2$ whereas Crystal B displays a featureless decrease in resistivity down to the onset of superconductivity.  Crystal A has a multiple step transition with a higher onset temperature and a lower, zero resistance temperature.  Crystal B has a sharp superconducting transition that, as will be shown below, does not significantly broaden in applied magnetic fields of up to 140 kOe.  These results are consistent with the evaluation of the amount of K in the sample shown in Fig. \ref{F2}; there can be a somewhat differing amount of K in different layers or different crystals of these compounds.
	
\begin{figure}
\begin{center}
\includegraphics[angle=0,width=120mm]{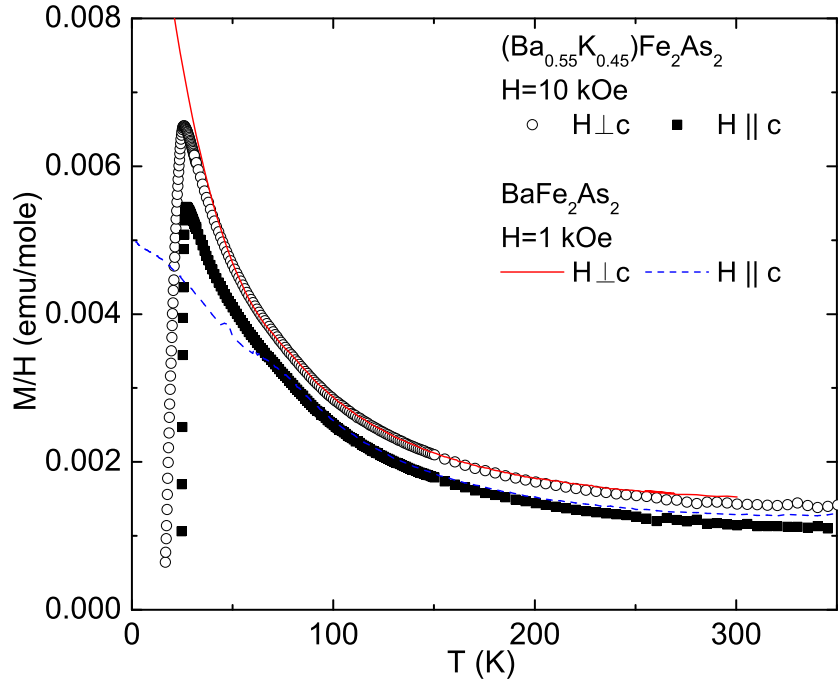}
\end{center}
\caption{(Color online) Temperature dependent, anisotropic magnetization divided by applied field for (Ba$_{0.55}$K$_{0.45}$)Fe$_2$As$_2$.  The dashed and solid lines are the data on BaFe$_2$As$_2$ from Fig. 5 for comparison.}\label{F10}
\end{figure}

Figure \ref{F10} presents the anisotropic $M/H$ data for (Ba$_{0.55}$K$_{0.45}$)Fe$_2$As$_2$.  The data for both directions of applied field are essentially featureless down to $T_c \sim 30$ K.  Below T ~ 30 K the data become strongly diamagnetic and will be discussed below. For comparison, the anisotropic $M/H$ data presented in Fig. \ref{F5} for BaFe$_2$As$_2$ are shown as dashed lines.  For $T > 90$ K the BaFe$_2$As$_2$ and (Ba$_{0.55}$K$_{0.45}$)Fe$_2$As$_2$ data are essentially identical, down to their slight, high temperature anisotropy.  The primary differences between the BaFe$_2$As$_2$ and (Ba$_{0.55}$K$_{0.45}$)Fe$_2$As$_2$ data sets are: (i) the effect of the 85 K phase transition in BaFe$_2$As$_2$ which causes the $H \| c$ data to break away from the corresponding data for (Ba$_{0.55}$K$_{0.45}$)Fe$_2$As$_2$, and (ii) the advent of superconductivity for $T < 30$ K.
	
\begin{figure}
\begin{center}
\includegraphics[angle=0,width=120mm]{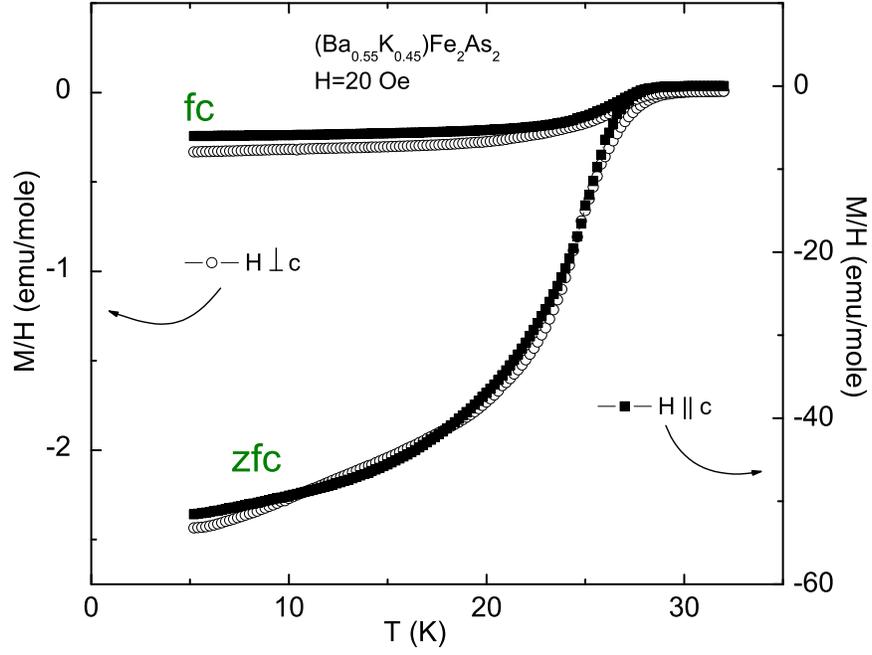}
\end{center}
\caption{Anisotropic, low field $M/H$ data for a (Ba$_{0.55}$K$_{0.45}$)Fe$_2$As$_2$ single crystal plate.  Field cooled, warming (fc) and zero field cooled, warming (zfc) data are shown for both orientations of the sample.  For $H \perp c$ the demagnetization factor will be the closest to zero.}\label{F11}
\end{figure}

The low field $M/H$ data for (Ba$_{0.55}$K$_{0.45}$)Fe$_2$As$_2$ are shown in Fig. \ref{F11} for magnetic field applied along the $c$-axis (perpendicular to the thin, plate-like sample) and perpendicular to the $c$-axis (along the surface of the thin, plate-like sample).  The temperature dependence of the field-cooled-warming and zero-field-cooled-warming data for the two field orientations are similar and an estimate of the superconducting fraction from the ZFC data for the field along the surface of the plate (the low demagnetization factor direction) is $\sim 50$\% of $1/4 \pi$. For $T < 4$ K there is a second diamagnetic step associated with the superconducting transition in the small amount of elemental Sn on the surface of the sample. (For clarity, these data are not shown.)
	
\begin{figure}
\begin{center}
\includegraphics[angle=0,width=120mm]{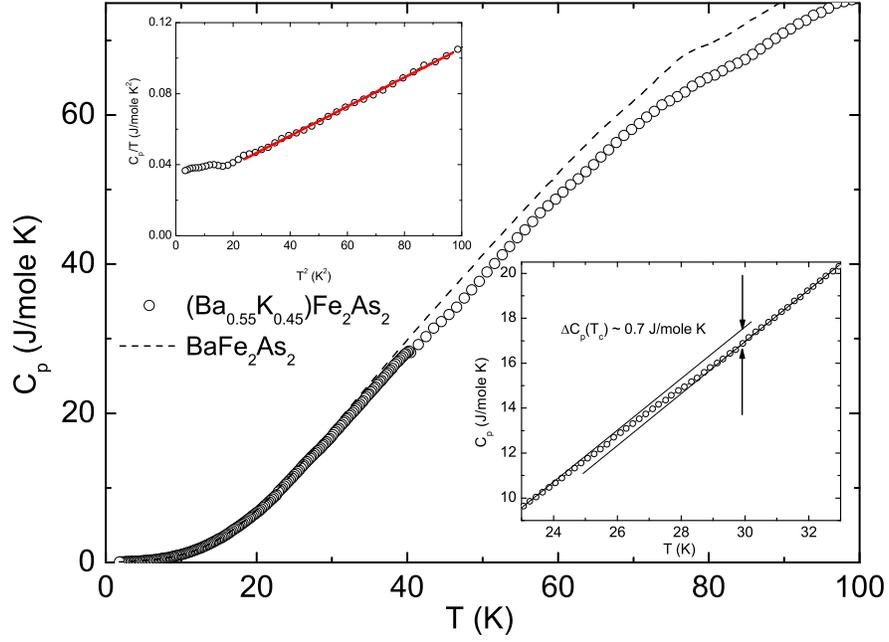}
\end{center}
\caption{(Color online) Temperature dependent specific heat of (Ba$_{0.55}$K$_{0.45}$)Fe$_2$As$_2$.  Upper inset:  $C_p/T$ as a function of $T^2$ for low temperature data.  Lower inset: expanded view near $T_c \sim 30$ K with linear extrapolations of $T > 30$ K and $T < 24$ K extended past $T_c$ to more clearly show the jump in $C_p$ at $T_c$.}\label{F12}
\end{figure}

The temperature dependent specific heat data for (Ba$_{0.55}$K$_{0.45}$)Fe$_2$As$_2$ are presented in Fig. \ref{F12}.  The data for BaFe$_2$As$_2$ (Fig. \ref{F6}) are shown for comparison as a dotted line.  The broad feature associated with the 85 K structural transition in BaFe$_2$As$_2$ is reduced, although there appears to be some remnant of it in the (Ba$_{0.55}$K$_{0.45}$)Fe$_2$As$_2$ data.  The low temperature specific heat data from (Ba$_{0.55}$K$_{0.45}$)Fe$_2$As$_2$ (upper inset) can be fit to the $C_p = \gamma T + \beta T^3$ power law for 100 K$^2 > T^2 > 20$ K$^2$ with $\gamma = 23$ mJ/mole K$^2$ and $\beta = 0.83$ mJ/mole K$^4$  from which a $\Theta_D$ value of 230 K can be inferred.  The small structure near $T^2 \sim 16$ K may well be associated with some residual Sn flux on the sample.  Regardless of this, it is worth noting that the linear, or electronic, specific heat term is clearly not going to zero as temperature approaches 0 K.  Indeed the estimated, low temperature electronic specific heat is not greatly different from that of the undoped BaFe$_2$As$_2$ shown in Fig. \ref{F6}.
	
The lower inset of Fig. \ref{F12} displays an expanded scale for $T \sim T_c$.  The data for $T > 30$ K and the data for $T < 24$ K are extended past $T_c$ so as to emphasize the slight, but detectable "jump" in specific heat of ~ 0.7 J/mole K associated with the onset of superconductivity.  In order to estimate the ratio $\Delta C_p/\gamma T_c$ the electronic specific heat in the normal state of (Ba$_{0.55}$K$_{0.45}$)Fe$_2$As$_2$ is needed.
Unfortunately, as will be shown below, superconductivity is not suppressed significantly by an applied magnetic field of 140 kOe.
Using the low temperature gamma values for BaFe$_2$As$_2$ and (Ba$_{0.55}$K$_{0.45}$)Fe$_2$As$_2$ as well as $T_c \sim 30$ K the ratio of $\Delta C_p/\gamma T_c$ is found to be 0.63 and 1.01 respectively. We can treat these data in an extreme limit by assuming that about half of the sample (layers) was not superconducting. This would explain the finite $\gamma$ at low temperatures and would increase the $\Delta C_p/\gamma T_c$ value of $\sim 0.63$ to $\sim 1.2$. Further measurements will be needed to refine this value more fully. To within the accuracy or our current measurements and estimates it is reasonable to evaluate $\Delta C_p/\gamma T_c \sim 1$.
	
\begin{figure}
\begin{center}
\includegraphics[angle=0,width=120mm]{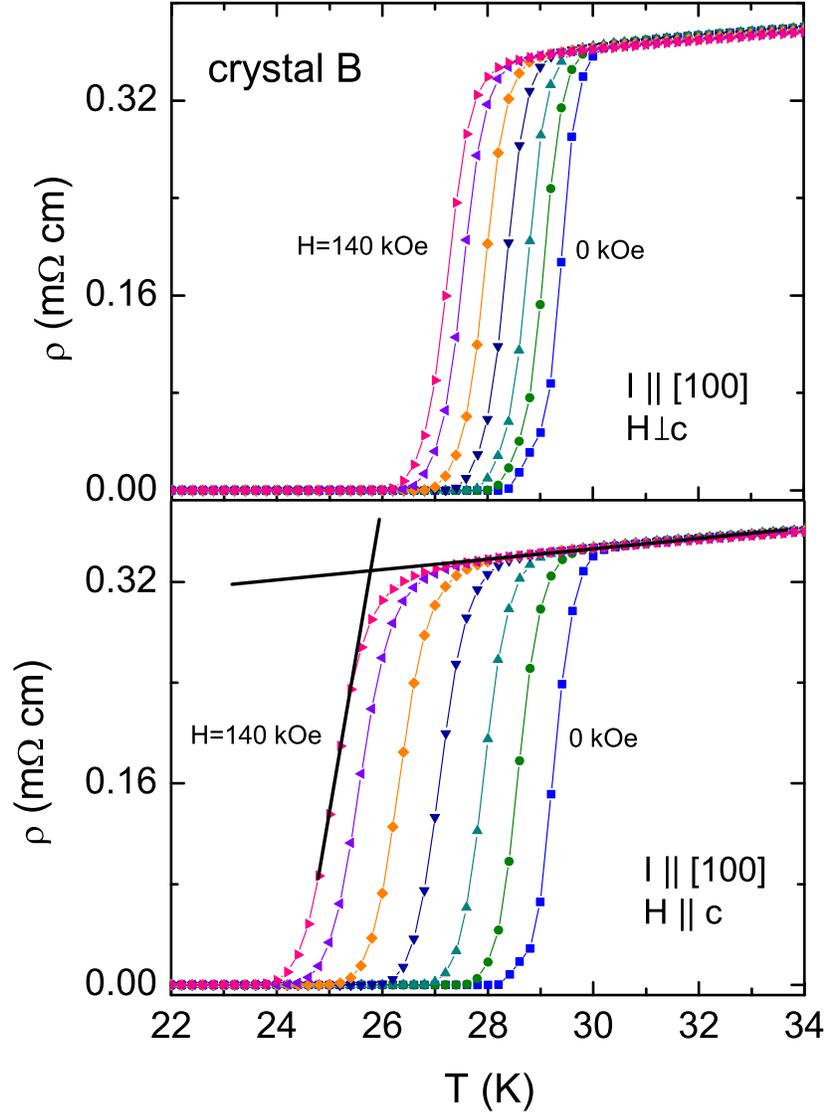}
\end{center}
\caption{(Color online) Temperature dependence of the electrical resistivity of (Ba$_{0.55}$K$_{0.45}$)Fe$_2$As$_2$ near $T_c$ in constant applied magnetic fields of up to 140 kOe for $H \perp c$ (upper panel) and $H \| c$ (lower panel).  Data (from higher transition temperature to lower) are for applied magnetic fields of 0, 10, 30, 60, 90, 120, 140 kOe respectively.  The lines drawn on the 140 kOe data in the lower panel illustrate the "onset" criterion that has been used to determine $T_c$ for each field.}\label{F13}
\end{figure}

\begin{figure}
\begin{center}
\includegraphics[angle=0,width=120mm]{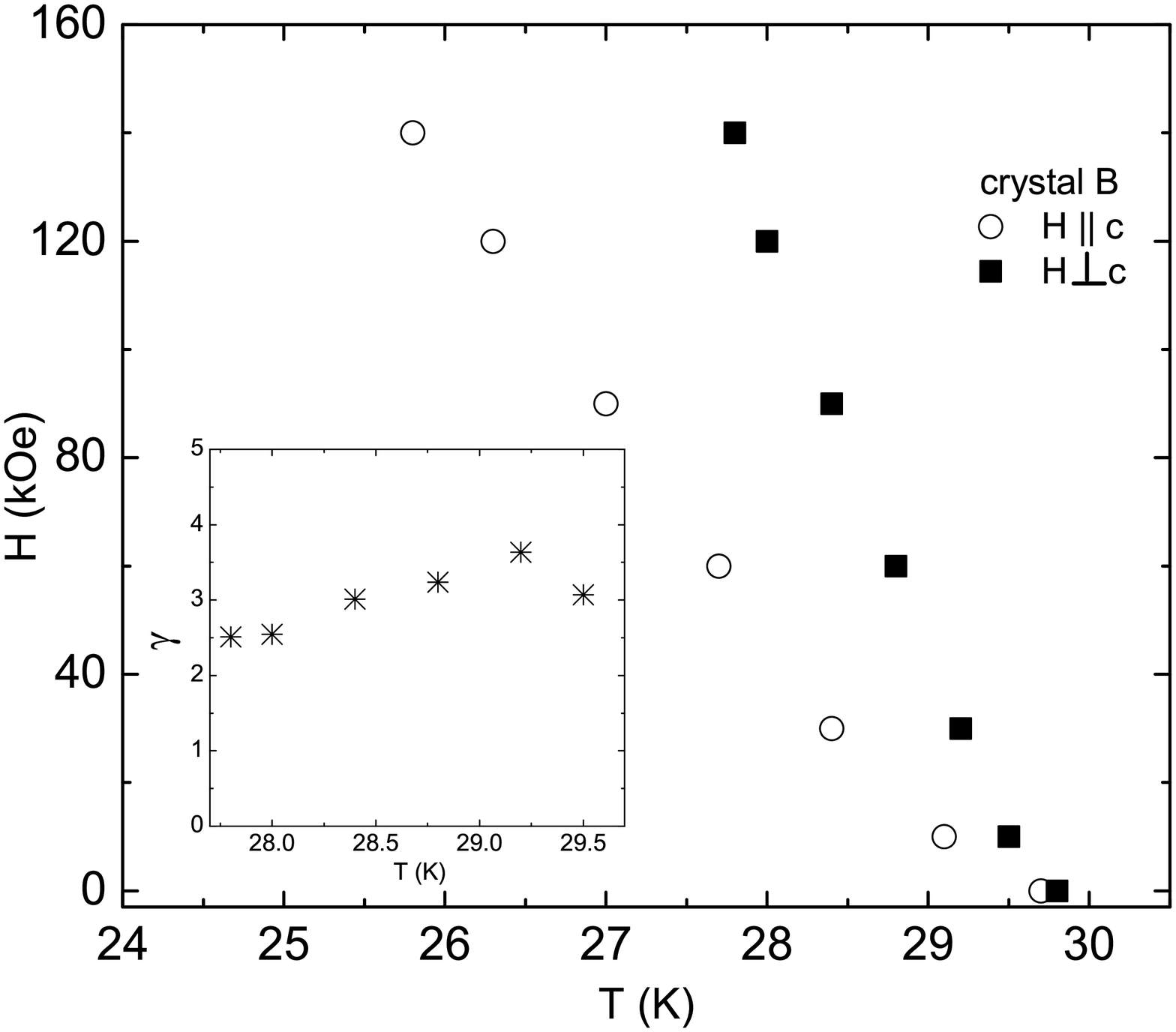}
\end{center}
\caption{Anisotropic $H_{c2}(T)$ plot for $H \le 140$ kOe.  Inset:  $\gamma = H_{c2}^{\perp c} / H_{c2}^{\| c}$ as a function of temperature just below $T_c$.}\label{F14}
\end{figure}

The anisotropy of the field dependence of the electrical resistivity, and suppression of superconductivity in (Ba$_{0.55}$K$_{0.45}$)Fe$_2$As$_2$ (crystal B) are shown in Fig. \ref{F13}.  The normal state resistivity is insensitive to applied fields as large as 140 kOe and the applied field suppresses the superconducting transition without any significant broadening of the transition.  Using an "onset" criterion for the determination of $T_c(H)$, $H_{c2}(T)$ curves were determined for both directions of applied field (Fig. \ref{F14}).  Similar suppressions and anisotropy were found for crystal A.
	
The ratio of the $H_{c2}$ values for the two characteristic directions can be defined as $\gamma = H_{c2}^{\perp c} / H_{c2}^{\| c}$ and is plotted in the inset to Fig. \ref{F14}.  $\gamma$ varies between 3.5 and 2.5 in the very limited range below $T_c$ we are able to measure (being limited by a 140 kOe maximum applied field and the extremely high slope of $H_{c2}^{\perp c}$).

\section{Analysis and Discussion}
Growth of single crystals of BaFe$_2$As$_2$ and (Ba$_{0.55}$K$_{0.45}$)Fe$_2$As$_2$ out of excess Sn flux provides direct access to single crystals of  the recently discovered \cite{sec_ger} family of intermetallic iron-arsenide superconductors.  Although the single crystals of BaFe$_2$As$_2$ manifest a transition near $\sim$ 85 K rather than the 140 K reported for polycrystalline samples, \cite{fst_ger} single crystal diffraction data indicate that the transition is the same type of tetragonal to orthorhombic phase transition.
	
The changes in the resistivity as well as the Hall resistivity of the BaFe$_2$As$_2$ crystals upon cooling through 85 K (Fig. \ref{F4}) suggest a decrease in the density of states at the Fermi surface associated with the structural phase transition to the lower temperature, orthorhombic phase.  It is not immediately clear why our basal plane resistivity manifests an approximately 50\% increase upon cooling through this phase transition while the polycrystalline data shows an approximately 80\% decrease in resistivity associated with the orthorhombic state. \cite{fst_ger}  This may simply be an intrinsic effect associated with the anisotropy of the electrical resistivity or it may be an extrinsic effect associated with the possible incorporation of $\sim 1$\% of Sn into the sample.
	
\begin{figure}
\begin{center}
\includegraphics[angle=0,width=120mm]{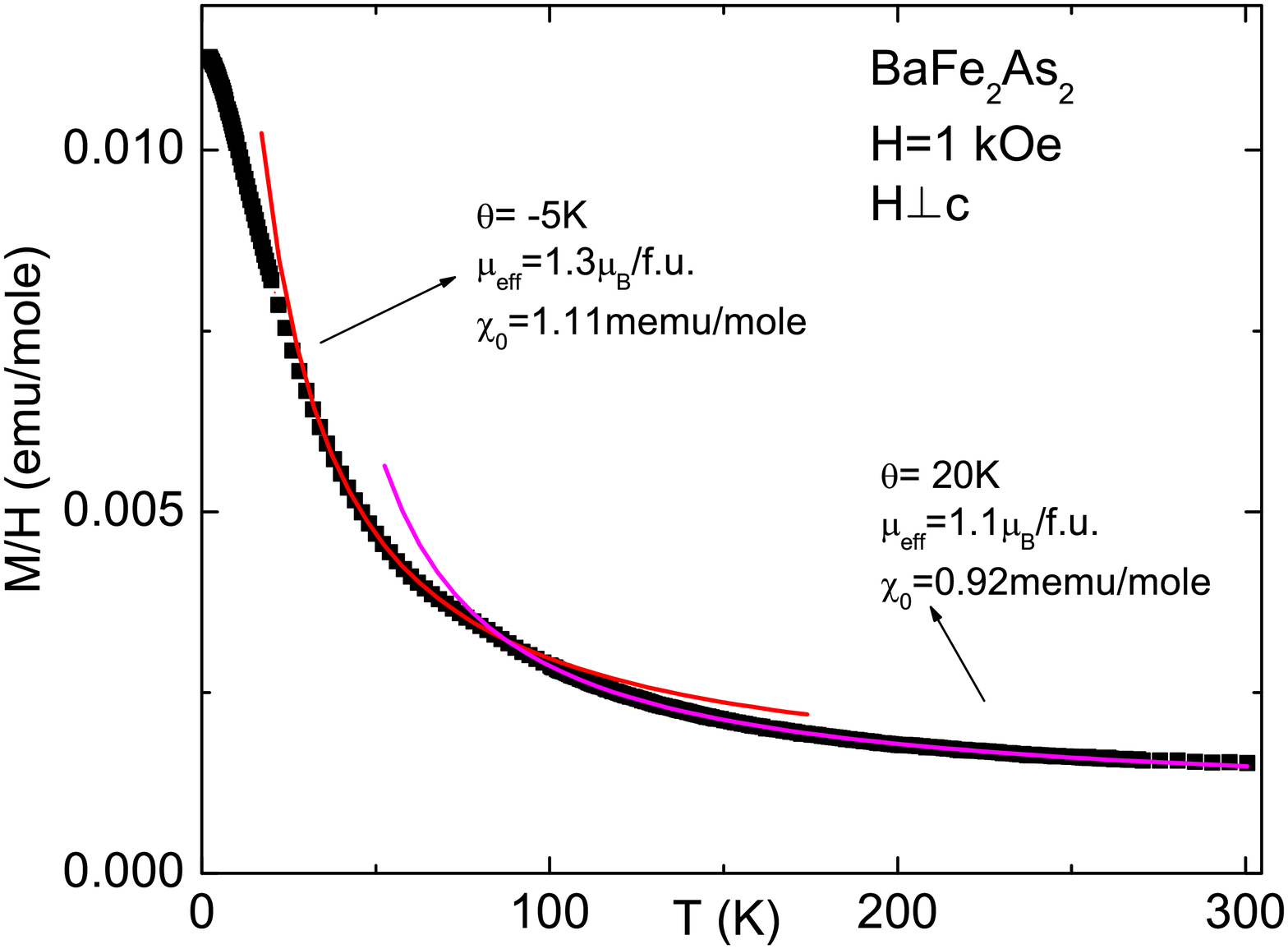}
\end{center}
\caption{(Color online) Curie-Weiss fits to high and low temperature (above and below $T \sim 85$ K) $M/H$ data for BaFe$_2$As$_2$ with $H \perp c$.}\label{F15}
\end{figure}

The magnetic susceptibility of  BaFe$_2$As$_2$ as well as (Ba$_{0.55}$K$_{0.45}$)Fe$_2$As$_2$ can be analyzed in terms of a Curie-Weiss law with an additional, temperature independent term to account for the sum of core diamagnetism, Pauli paramagnetism and Landau diamagnetism:  $\chi = C/(T - \theta) + \chi_0$.  Figure \ref{F15} shows the temperature dependent $M/H$ data for BaFe$_2$As$_2$ for $H \perp c$.  The high and low temperature parts (above and below $T \sim 85$ K) of this curve can be well fit by a Curie-Weiss form with similar values of both $\chi_0$ and $\mu_{eff}$.  The primary effect of the structural phase transition appears to be to cause a change in the sign of the paramagnetic $\theta$ from +20 K in the high temperature region to -5 K in the low temperature region. This could be consistent with a change in the nature of the in-plane magnetic interactions; microscopic measurements will be needed to shed further light on this.  Given the similarity between the high temperature $M/H$ data for BaFe$_2$As$_2$ and (Ba$_{0.55}$K$_{0.45}$)Fe$_2$As$_2$ shown in Fig. \ref{F10}, it is is not surprising that the Curie-Weiss fit to the (Ba$_{0.55}$K$_{0.45}$)Fe$_2$As$_2$ data is similar to that for the BaFe$_2$As$_2$.  Within the accuracy of these fits (given their limited temperature range) these data are consistent with $\sim 1.2 \mu_B$/formula unit for all temperature ranges and compositions.
	
The $H_{c2}(T)$ data shown in Fig. \ref{F14} clearly indicate that (Ba$_{0.55}$K$_{0.45}$)Fe$_2$As$_2$ will have a substantial $H_{c2}(0)$ value.  Depending on the model or extrapolation chosen, $H_{c2}(0)$ for $H \| c$ can be expected to be anywhere between 750 kOe and 1100 kOe.  This lower value only slightly exceeds the Clogston paramagnetic limit. \cite{clo} If the anisotropy seen close to $T_c$ (Fig. \ref{F14}) remains unchanged, $H_{c2}(0)$ for $H \perp c$ will be even larger.  These are very long extrapolations and these values only serve to indicate that (Ba$_{0.55}$K$_{0.45}$)Fe$_2$As$_2$ will be a very high field superconductor.

\section{Summary}
Anisotropic thermodynamic and transport measurements, as well as crystallographic measurements, have been performed on sizable single crystals of BaFe$_2$As$_2$ and (Ba$_{0.55}$K$_{0.45}$)Fe$_2$As$_2$ grown out of Sn flux. The crystals of both composition have small but detectable ($\sim 1.0$\% and $\sim 0.66$\% respectively) amounts of Sn incorporated into the bulk and in BaFe$_2$As$_2$ this is correlated with a reduction of the tetragonal to orthorhombic, structural phase transition temperature to $\sim 85$ K and changes in the temperature dependencies of the electrical resistivity and magnetic susceptibility.  On the other hand the (Ba$_{0.55}$K$_{0.45}$)Fe$_2$As$_2$ manifest clear, bulk superconductivity below $T \sim 30$ K.
Both compounds manifest a strong, temperature dependent, paramagnetic response with an effective moment of $\sim 1.2 \mu_B$/formula unit, independent of doping or structure.  So, although Sn clearly effects the very sensitive structural phase transition in BaFe$_2$As$_2$, it does not significantly effect the high temperature, superconducting phase transition in (Ba$_{0.55}$K$_{0.45}$)Fe$_2$As$_2$.

The superconducting transition in (Ba$_{0.55}$K$_{0.45}$)Fe$_2$As$_2$ can be clearly detected in electrical resistivity and magnetization measurements and can also be observed, in the specific heat data.  Based on the anomaly in $C_p(T)$ we can estimate $\Delta C_p/\gamma T_c \sim 1$.  The upper superconducting critical field is clearly anisotropic with $\gamma = H_{c2}^{\perp c} / H_{c2}^{\| c}$ varying between 3.5 and 2.5 for the first two degrees below $T_c$.  (Ba$_{0.55}$K$_{0.45}$)Fe$_2$As$_2$ will clearly have an exceptionally high $H_{c2}(0)$ with estimates of 750 kOe or higher being easily justifiable.

% tables should appear as floats within the text
%
% Here is an example of the general form of a table:
% Fill in the caption in the braces of the \caption{} command. Put the label
% that you will use with \ref{} command in the braces of the \label{} command.
% Insert the column specifiers (l, r, c, d, etc.) in the empty braces of the
% \begin{tabular}{} command.
% The ruledtabular enviroment adds doubled rules to table and sets a
% reasonable default table settings.
% Use the table* environment to get a full-width table in two-column
% Add \usepackage{longtable} and the longtable (or longtable*}
% environment for nicely formatted long tables. Or use the the [H]
% placement option to break a long table (with less control than
% in longtable).
% \begin{table}%[H] add [H] placement to break table across pages
% \caption{\label{}}
% \begin{ruledtabular}
% \begin{tabular}{}
% Lines of table here ending with \\
% \end{tabular}
% \end{ruledtabular}
% \end{table}

% Surround table environment with turnpage environment for landscape
% table
% \begin{turnpage}
% \begin{table}
% \caption{\label{}}
% \begin{ruledtabular}
% \begin{tabular}{}
% \end{tabular}
% \end{ruledtabular}
% \end{table}
% \end{turnpage}

% Specify following sections are appendices. Use \appendix* if there
% only one appendix.
%\appendix
%\section{}

\begin{acknowledgments}
Work at the Ames Laboratory was supported by the US Department of Energy - Basic Energy Sciences under Contract No. DE-AC02-07CH11358.  The authors would like to acknowledge motivations two and four and useful discussions with V. G. Kogan, R. W. McCallum, K. Dennis, E. D. Mun, M. Tillman, R. Prozorov, G. D. Samolyuk, M. Tanatar, J. Q. Yan, M. Lampe, and F. Laabs.

\end{acknowledgments}

\clearpage

\begin{table}
\caption{WDS elemental analysis (in atomic \%) for (Ba$_{0.55}$K$_{0.45}$)Fe$_2$As$_2$ single crystal shown in Fig. \ref{F2}.\label{T1}}
%\begin{ruledtabular}
\begin{tabular}{c c c c c c c}
\hline\hline
Point & ~~As~~ & ~~Sn~~ & ~~K~~ & ~~Fe~~ & ~~Ba~~ & K/(K+Ba) \\
\hline
1 & 37.6 & 0.53 & 10.2 & 41.7 & 9.9 & 0.51 \\
2 & 38.5 & 0.74 & 8.1 & 40.3 & 12.3 & 0.40 \\
3 & 38.3 & 0.89 & 6.8 & 42.1 & 12.0 & 0.36 \\
4 & 38.1 & 0.93 & 7.0 & 41.5 & 12.6 & 0.36 \\
5 & 38.4 & 0.48 & 10.3 & 40.6 & 10.3 & 0.50 \\
6 & 38.3 & 0.48 & 10.7 & 40.9 & 9.7 & 0.53 \\
7 & 38.5 & 0.74 & 8.8 & 41.4 & 10.6 & 0.45 \\
8 & 38.2 & 0.71 & 9.4 & 41.4 & 10.2 & 0.48 \\
\hline
\end{tabular}
% \end{ruledtabular}
\end{table}

\clearpage

\clearpage

\clearpage

\clearpage

\clearpage

\clearpage

\clearpage

\clearpage

\clearpage

\clearpage

\clearpage

\clearpage

\clearpage

\clearpage

\clearpage

\begin{thebibliography}{99}

\bibitem{jap_dis}  Y. Kamihara, T. Watanabe, M. Hirano, H Hosono, Journal of the American Chemical Society  {\bf 130}, 3296 (2008).

\bibitem{jap_pre}  H. Takahashi, K. Igawa, K. Arii, Y. Kamihara, M. Hirano, H. Hosono, Nature (London)  {\bf 453}, 376 (2008).

\bibitem{fst_sm}  X. H. Chen, T. Wu, G. Wu, R. H. Liu, H. Chen, D. F. Fang, Nature (London)  {\bf 453}, 761 (2008).

\bibitem{pro_po}  R. Prozorov, M. E. Tillman, E. D, Mun, P. C. Canfield, cond-mat/08052783, unpublished.

\bibitem{karp}  N. D. Zhigadlo, S. Katrych, Z. Bukowski, J. Karpinski cond-mat/08060337, unpublished.

\bibitem{sec_ger}  M. Rotter, M. Tegel, D. Johrendt, cond-mat/08054630, unpublished.

\bibitem{fst_ger}  M. Rotter, M. Tegel, D. Johrendt, I. Schellenberg, W. Hermes, R. Poettgen, cond-mat/08054021, unpublished.

\bibitem{can_fi}  P. C. Canfield, Z.  Fisk, Phil. Mag. B  {\bf 65}, 1117 (1992).

\bibitem{clo}  A. M. Clogston, Phys. Rev. Lett.  {\bf 9}, 266 (1962).

\end{thebibliography}
\end{document}